# Broadband SNAIL parametric amplifier with microstrip impedance transformer

D. Ezenkova,[1,2,a)] D. Moskalev,[1,a)] N. Smirnov,[1] A. Ivanov,[1] A. Matanin,[1]
V. Polozov,[1] V. Echeistov,[1] E. Malevannaya,[1] A. Samoilov,[1] E. Zikiy[1] and
I. Rodionov[1,2,b)]

[1]*FMN Laboratory, Bauman Moscow State Technical University, Moscow, 105005, Russia*

[2]*Dukhov Automatics Research Institute, VNIIA, Moscow, 127030, Russia*

Josephson parametric amplifiers have emerged as a promising platform for quantum information processing and squeezed quantum states generation. Travelling wave and impedance-matched parametric amplifiers provide broad bandwidth for high-fidelity single-shot readout of multiple qubit superconducting circuits. Here, we present a quantum-limited 3-wave-mixing parametric amplifier based on superconducting nonlinear asymmetric inductive elements (SNAILs), whose useful bandwidth is enhanced with an on-chip two-section impedance-matching circuit based on microstrip transmission lines. The amplifier dynamic range is increased using an array of sixty-seven SNAILs with 268 Josephson junctions, forming a nonlinear quarter-wave resonator. Operating in a current-pumped mode, we experimentally demonstrate an average gain of 17 dB across 300 MHz bandwidth, along with an average saturation power of –100 dBm, which can go as high as −97 dBm with quantum-limited noise performance. Moreover, the amplifier can be fabricated using a simple technology with just a one e-beam lithography step. Its central frequency is tuned over a several hundred megahertz, which in turn broadens the effective operational bandwidth to around 1.5 GHz.

Parametric amplifiers have become key components in quantum information processing due to their near quantum-limited noise performance. Those devices are used as the first stage of the low-noise amplification schemes defining primarily the overall system noise and signal-to-noise ratio [1]. Broadband Josephson parametric amplifiers are the only tools to perform single-photon power measurements of both microwave [2-4] and optical [5] signals, quantum metrology [6], and dark matter search [7]. In quantum computing it has been used for high-fidelity single-shot readout of superconducting qubits [8-12], real-time quantum feedback control [13-15], and generate squeezed quantum states [16, 17].

There are several types of parametric amplifiers that have been demonstrated in the last few years for multiplexed qubits readout [18-20]. The first one is an impedance-transformed Josephson parametric amplifier (IMPA). It allows to increase gain-bandwidth product due to impedance-matching of the amplifier to its input. This idea was first demonstrated using Klopfenstein taper structure to make a stronger coupling between JPA and its environment [21, 22]. Later, several scientific groups [23–25] proposed using a simpler two–section impedance transformers based on coplanar waveguides to broaden amplifier bandwidth. A maximum gain of 15-20 dB, a bandwidth in the range of 150-600 MHz, a saturation power of -120…-100 dBm, and a noise near the quantum limit have been demonstrated for these devices. However, the saturation power of those devices is limited as its parametric interaction is described by Kerr nonlinearity. Moreover, these devices require complex multi-stage fabrication process.

In order to enhance both bandwidth and dynamic range, the Josephson traveling-wave parametric amplifier (JTWPA) was introduced. These devices are based on a nonlinear transmission line demonstrating a bandwidth up to several gigahertz at 10-20 dB gain, which allow reading out dozens of superconducting qubits in a single shot [26–28]. Usually, it requires fabrication of several thousands of nearly identical nonlinear elements (Josephson junctions and parallel plate capacitors), demanding sophisticated fabrication facilities which are not widely available. Furthermore, the noise temperature of such devices is 2-4

---

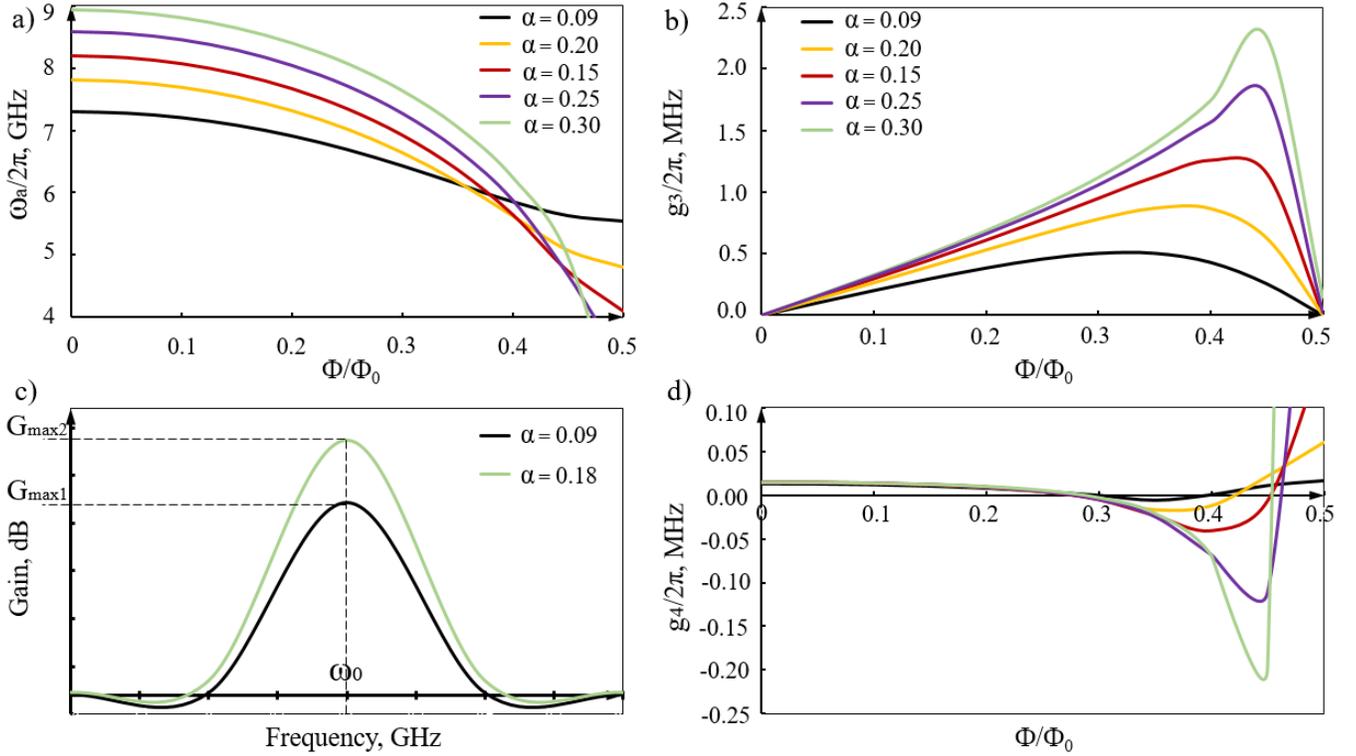

FIG. 1 SNAIL Nonlinear characteristics. (a) Tunable range dependency of the SNAIL resonance frequency versus normalized magnetic flux for different SNAIL asymmetries α at device capacitance C = 30fF. (b) Third-order nonlinearity ($g_3$) SNAIL asymmetry. (c) JPA gain profile for α = 0.09 (black) and α = 0.18 (green) at central resonance frequency $\omega_a/2\pi$ = 6.4 GHz. (d) Forth-order nonlinearity ($g_4$) and Kerr-free area dependence of SNAIL asymmetry.

times higher the quantum limit due to the losses in the circuit [29]. Recent advances in this direction are associated with the development of JTWPA based on superconducting nonlinear asymmetric inductive elements [30]. This approach ensures eliminating the gap in the dispersion relation, but the noise temperature is still too high.

In this work, we present a quantum-limited 3-wave mixing impedance-matched JPA consisting of SNAIL array with on-chip two-section microstrip impedance transformer. It operates in reflection mode when the input signal reflects off, generating the amplified output signal with the gain of more than 17 dB and idler tone. To improve the dynamic range of our degenerate parametric amplifier we use the SNAILs, which provide the flexibility in optimizing a 3-wave-mixing amplification process, while simultaneously minimizing a 4-wave-mixing Kerr nonlinearity suspected to cause amplifier saturation [31–33]. The IMPA saturation power scales as $I_c^2/Q^3$ (where $I_c$ is the critical current of Josephson junctions and Q is the coupled (external) Q-factor allowing to increase the saturation power without degradation other important parameters. Based on this, we have engineered the IMPA with 1 dB compression power P$_{-1dB}$ ∈ [−97, −100] dBm at 17 dB gain and quantum-limited noise performance.

Resonance frequency of the IMPA is defined as $\omega_0 = 1/\sqrt{L_{array}C}$, where $L_{array}$ – SNAIL array inductance which can be defined as in (1) and C – equivalent circuit capacitance [31].

$$L_{array} = ML_s \qquad (1)$$

$$L_s(\varphi_{ext}) = L_J/c_2(\varphi_{ext}) \qquad (2)$$

where M – number of SNAILs in array, $L_j$ – Josephson inductance of large junctions, $c_2 = \alpha \cos\varphi_{min} + \frac{1}{3}\cos\left(\frac{\varphi_{min}-\varphi_{ext}}{3}\right)$ – flux-tunable constant, α – cell asymmetry, $\varphi_{ext} = 2\pi\Phi/\Phi_0$ – magnetic flux quantum.

IMPA tunable range is directly defined by SNAIL asymmetry $\alpha$, which depends on the junctions critical current ratio. Maximum asymmetry value defined as 1/n, n – number of large Josephson junctions in the loop. Decreasing SNAIL asymmetry leads to the tunable range

a) D. Ezenkova and D. Moskalev contributed equally to this work

b) Electronic mail: irodionov@bmstu.ru.

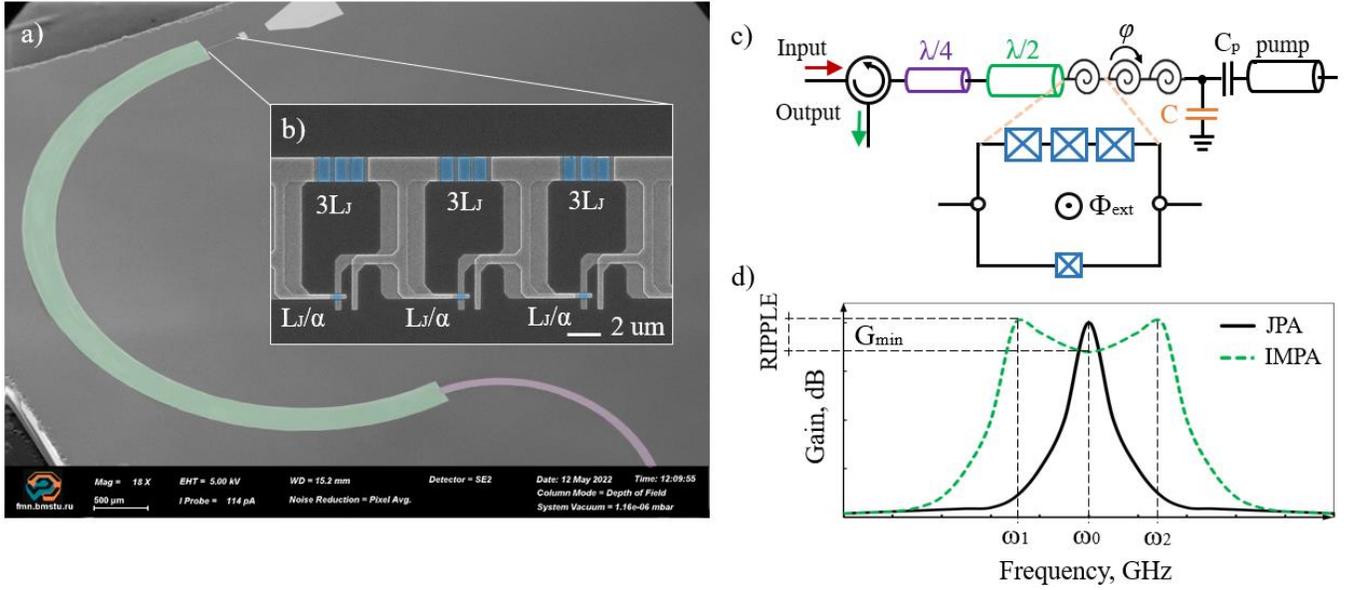

FIG 2. Experimental device. (a) Micrograph of the fabricated device with the series of λ/4 microstrip transformer (purple), λ/2 microstrip transformer (green), SNAIL array and 30 fF capacitor (orange). (b) Higher-magnification micrograph of SNAIL, which consists of an array of 3 large Josephson junctions (blue) in a loop with one smaller junction (blue). (c) Circuit representation of the device. (d) IMPA gain profile for n=2

of parametric amplifier decreasing [FIG. 1(a)]. Number of SNAILs M and asymmetry coefficient $\alpha$ characterize nonlinear processes in the device. SNAIL has both types of nonlinearities, where a third-order nonlinearity ($g_3$) depicts 3-wave-mixing process and maximum gain and a fourth-order nonlinearity or Kerr nonlinearity ($g_4$) describes 4-wave-mixing process and saturation power of parametric amplifier. The ratio between third- and fourth-order nonlinearities defines the IMPA working point.

FIG. 1(b) shows the dependency of the SNAIL third-order nonlinearity versus normalized magnetic flux for various asymmetry coefficients $\alpha$ at a fixed value of Josephson inductance $L_j$ and number of nonlinear elements M. One can clearly notice, it is non-zero in the whole range of magnetic flux values except $\Phi/\Phi_0 = 0$ and $\Phi/\Phi_0 = 0.5$ points, where 3-wave mixing term is forbidden. At small asymmetry values $\alpha=0.09$ (black) third-order nonlinearity $g_3$ has a flat profile, which can provide a stable 20 dB gain in the whole tunable range of a narrow-banded Josephson parametric amplifier. As $\alpha$ increases, the third-order nonlinearity value is increased providing higher gain at the central resonance frequency [see FIG. 1(c)]. For a greater part of the flux range the fourth-order nonlinearity $g_4$ is positive, but the SNAILs have a region with the negative values of $g_4$ (Kerr-free region), where $g_3$ is dominant and the three-wave mixing process takes place [see FIG. 1(d)].

For impedance-transformer modelling the negative-resistance prototypes method [34] was utilized. We used the $2^{nd}$ order prototype (n=2), which has the Chebyshev ripple in its passband [see FIG. 2(c)]. The gain is defined as the ratio of the reflected power, dissipated in the load resistance, to the power available from the generator. The key task of the modelling is to achieve a reasonable correlation between the prototype and the actual circuitry of parametric amplifier. It requires creating lumped-element circuit with the prototype gain and ripple as well as designed center-frequency, impedance and bandwidth.

Prototype parameters and parametric amplifier circuit characteristics defined by impedance or admittance slope parameter. An actual negative-resistance device (parametric amplifier) determine the minimum value of the slope parameter of the λ/2 resonator. For electrical prototype circuit proposed in FIG. 2 impedance of λ/4 and λ/2 transformers can be defined as:

$$Z_{\lambda/4}^2 = Z_0 R_0 \quad (3)$$

$$Z_{\lambda/2} = \pi/2b \quad (4)$$

$$Z_{JPA} = \sqrt{L_{array}/C} = 2x/\pi \quad (5)$$

---

a) D. Ezenkova and D. Moskalev contributed equally to this work

b) Electronic mail: irodionov@bmstu.ru.

where $R_0 = R/r_0$ – source impedance, $b = g_2/wR = \omega_0 C_{ptype}$ – admittance slope parameter, $x = g_1 R/w$ – impedance slope parameter. It's important condition that each transformers resonance frequency have to be equal to the parametric amplifier resonance frequency ($\omega_{\lambda/4} = \omega_{\lambda/2} = \omega_{JPA} = \omega_0$). The impedance-transformer provides an increase of the amplifier bandwidth lowering the quality factor of the IMPA resonance. The proposed method results in a multi-peak gain profile [see FIG. 2(d)]. The ripples of the gain defined by negative-resistance prototype parameters: number of sections, section impedance values and negative-resistance value. We first carefully match circuit design and prototype parameters, then simulate full IMPA circuit in Ansys HFSS.

FIG. 2(a, b) shows scanning electron microscopy images of the impedance-matched Josephson parametric amplifier and its equivalent circuit. It consists of an array of M SNAILs connected in series with 30 fF capacitor. Each SNAIL consists of an array of 3 large Josephson junctions (with Josephson inductance $L_J = 80$ pH) in a loop with one smaller junction (with inductance $L_J/\alpha$) which has an asymmetry $\alpha = 0.18$ [see FIG. 2(b)]. Impedance transformer is implemented by a series connection of λ/4 resonator with 87Ω characteristic impedance, lowering the quality factor of the JPA resonance, and λ/2 resonator with 59Ω characteristic impedance, which increases the amplification bandwidth. Both microstrip transmission lines centered at 6.4 GHz corresponding to JPA central resonance frequency ($\omega_{\lambda/4} = \omega_{\lambda/2} = \omega_{JPA} = 6.4$ GHz).

The IMPA fabrication includes a two-step process: (I) SNAIL and impedance transformer pattering with double-angle evaporation and lift-off, and (II) wafer backside metallization. The described above processes starts on a high-resistivity silicon substrate (525 μm thick). Prior to exposure the substrate is cleaned in Piranha solution at 80°C. Then, a bilayer mask is spin-coated on the substrate, which consists of 500 nm MMA (methyl methacrylate) and 150 nm AR-P (CSAR). Impedance transformer input line and SNAIL array are patterned using direct 50 kV e-beam lithography. Al/AlOx/Al Josephson junction and microstrip impedance transformer are e-beam shadow-evaporated in a single vacuum cycle. Lift-off is performed in a bath of N-methyl-2-pyrrolidone with sonication at 80°C for 3 hours and rinsed in IPA with ultrasonication. The microstrip transmission line are finally formed using substrate backside silver ground plane deposited.

Finally, we experimentally measured our amplifier in a dilution refrigerator with a base temperature near 10 mK. The cryogenic characterization was done in reflection mode with circulator connected in series. The flux bias of the SNAIL loops was controlled by superconducting coil. Here, we define the operational bandwidth as the range of frequencies where the gain is greater than 17 dB and noise temperature corresponds to the standard quantum limit. First, we determine the IMPA characteristic frequency by applying flux to the SNAIL loop through an external coil. As shown on the FIG. 3(a) the resonant frequency can be tuned from 4.9 GHz to 8.2 GHz. It is well known the IMPA gain profile depends on the components of cryogenic setup. In our model, we assume an ideal 50 Ω impedance connected to the amplifier circuit. The real measurement circuit includes non-ideal cables, circulators and wire-bond connections, which also affect the gain profile.

FIG. 3(b) shows the best gain profile obtained at central frequency $\omega_{JPA} = 6.4$ GHz with the bandwidth over 300 MHz. This profile was obtained at the pump power of $P_{pump} = -11,5$ dBm and DC coil current of 2 mA. Measured device operates in a three-wave mixing mode. The saturation power of the amplifier is defined as a value of the input signal power at which the gain is decreased by 1 dB. The saturation power was also measured at the center frequency $\omega_{JPA} = 6.4$ GHz. We measured the saturation power of −97…−100 dBm in the bandwidth of 300 MHz with the gain above 17 dB [see FIG. 3(b)].

For noise temperature measurement we used the method described in [21]. We calibrated the noise of the HEMT following our IMPA, and then calculate system noise using the SNR improvement method. The estimated IMPA noise temperature is consistent with near-quantum-limited operation [see FIG. 3(b)].

In summary, we have designed, fabricated, and characterized the impedance-matched Josephson parametric amplifier based on SNAIL array and microstrip impedance transformer connected in series. The proposed device does not require complicated multi-step process and can be fabricated with just a one single e-beam lithography. We demonstrated an average gain of 17 dB with a bandwidth over 300 MHz at central resonance frequency of 6.4 GHz, that corresponds to design frequency. The noise temperature was estimated to be close to standard quantum limit with the saturation power in the range of [−97, −100] dBm. We experimentally tested the proposed device for ultrahigh efficient multi-resonator quantum memory readout, where a broadband amplifier is required to perform quantum tomography of the quantum states storing process with a low number of photons [35]. Devices

a) D. Ezenkova and D. Moskalev contributed equally to this work

b) Electronic mail: irodionov@bmstu.ru.

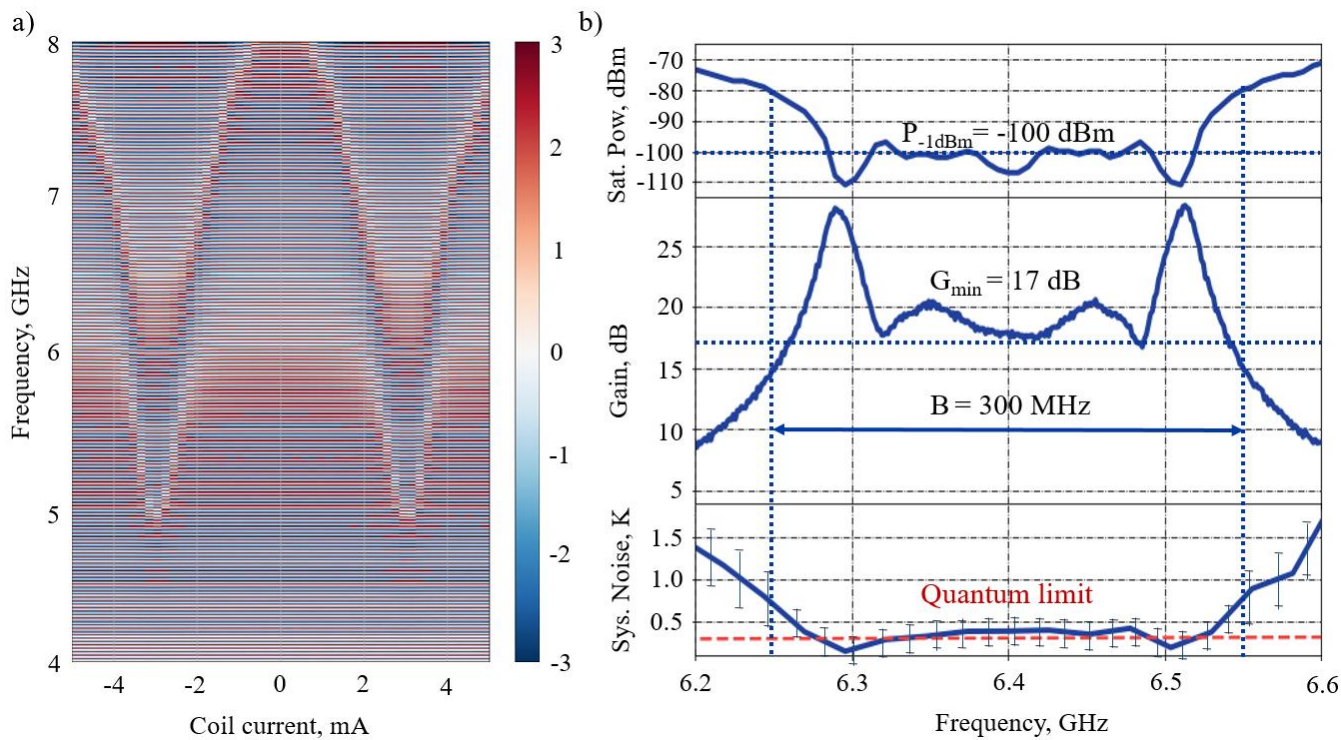

FIG. 3. Measured device performance operated at 10 mK. (a) Phase response of the IMPA vs magnetic flux bias controlled by superconducting coil. Input saturation power, gain and noise temperature of the IMPA (b). Here we demonstrate the amplifier performance at center frequency $\omega_{a/2\pi}$ = 6.4 GHz. The IMPA provides an input saturation power of -100 dBm with regions as high as -97 dBm. The device shows the gain above 17 dB and operational bandwidth of nearly 300 MHz. We define the quantum limit as one photon $\hbar\omega$ of total system noise at the input of the amplifier

were fabricated at the BMSTU Nanofabrication Facility (Functional Micro/Nanosystems, FMN REC, ID 74300).

## AUTHOR DECLARATIONS
### Conflict of Interest
The author have no conflicts to disclose.

### Author Contributions

**Daria Ezenkova:** Conceptualization (equal); Formal analysis (lead); Methodology (equal); Investigation (lead); Writing – original draft (lead). **Dmitry Moskalev:** Conceptualization (equal); Formal analysis (lead); Methodology (equal); Investigation (lead); Writing – review and editing (equal). **Nikita Smirnov:** Formal analysis (equal); Investigation (equal); Writing – review and editing (equal). **Anton Ivanov:** Formal analysis (equal); Investigation (equal); Writing – review and editing (equal). **Alexey Matanin:** Formal analysis (supporting); Investigation (supporting); Writing – review and editing (supporting). **Victor Polozov:** Investigation (supporting); Writing – review and editing (supporting). **Vladimir Echeistov:** Formal analysis (supporting); Investigation (supporting). **Elizaveta Malevannaya:** Formal analysis (supporting); Investigation (supporting). **Andrey Samoilov:** Formal analysis (supporting); Investigation (supporting); Writing – review and editing (supporting). **Evgeniy Zikiy:** Formal analysis (supporting); Investigation (supporting). **Ilya Rodionov:** Project administration (lead); Conceptualization (lead); Formal analysis (equal); Supervision (lead); Writing – review and editing (equal).

## DATA AVAILABILITY
The data that support the findings of this study are available from the corresponding author upon reasonable request.

---

a) D. Ezenkova and D. Moskalev contributed equally to this work

b) Electronic mail: irodionov@bmstu.ru.